\documentclass[aps,twocolumn,floatfix]{revtex4}
\usepackage{epsfig}
\usepackage{graphicx}
\usepackage{dcolumn}
\usepackage{natbib}
\usepackage{amssymb}

\def\dj{\hbox{d\kern-0,347em \vrule width0,3em height1,252ex
depth-1,21ex \kern0,051em}}
 
\begin{document}

\title{Effect of Disorder and Notches on Crack Roughness}

\author{Phani K.V.V. Nukala}
\affiliation{Computer Science and Mathematics Division, 
Oak Ridge National Laboratory, Oak Ridge, TN 37831-6164, USA}
\author{Stefano Zapperi}
\affiliation{CNR-INFM, SMC, Dipartimento di Fisica,
Sapienza --- Universit\`a di Roma, P.le A. Moro 2, 00185 Roma, Italy}
\affiliation{ISI Foundation, Viale S. Severo 65, 10133 Torino, Italy}
\author{Mikko J. Alava}
\affiliation{Laboratory of Physics, Helsinki University of Technology, 
FIN-02015 HUT, Finland}
\author{Sr{\dj}an \v{S}imunovi\'{c}}
\affiliation{Computer Science and Mathematics Division, 
Oak Ridge National Laboratory, Oak Ridge, TN~37831-6164, USA}
 
\begin{abstract}
We analyze the effect of disorder and notches on crack roughness in two dimensions. Our simulation 
results based on large system sizes and extensive statistical sampling indicate that 
the crack surface exhibits a universal local roughness of $\zeta_{loc} = 0.71$ and 
is independent of the initial notch size and disorder in breaking thresholds. The global 
roughness exponent scales as $\zeta = 0.87$ and is also independent of 
material disorder. Furthermore, we note that the 
statistical distribution of crack profile height fluctuations is also 
independent of material disorder and is described by a Gaussian 
distribution, albeit deviations are observed in the tails.
\end{abstract}
%\PACS{46.50.+a, 64.60.Ak}
\maketitle

\section{Introduction}

The statistical properties of fracture in disordered media are
interesting for theoretical reasons and practical applications \cite{breakdown,alava06}. 
An important theoretical issue is represented by the scaling of crack surfaces.
Experiments on several materials 
under different loading conditions have shown that the fracture surface is
self-affine \cite{man} and, in three dimensions, the out of plane roughness exponent 
displays a universal value of $\zeta \simeq 0.8$ irrespective of the
material studied \cite{bouch}. The scaling regime is sometimes quite impressive,
spanning five decades in metallic alloys \cite{bouch}. 
In particular, experiments have been
done in metals \cite{metals}, glass \cite{glass}, rocks \cite{rocks}  
and ceramics \cite{cera} covering both ductile and brittle
materials. Later on, a smaller exponent $\zeta=0.4-0.6$ was observed at smaller 
length scales. It was conjectured that 
crack roughness displays a universal value of $\zeta \simeq 0.8$ only at larger scales 
and at higher crack speeds, whereas another roughness exponent in the range of 
$0.4-0.6$ is observed at smaller length scales under quasi-static or slow crack 
propagation \cite{bouch}. It was recently shown that the short-scale 
value is not present in silica glass, even when cracks
move at extremely low velocities \cite{ponson06}. In addition,
in granite and sandstone, one only measures a value 
of $0.45$ even at high velocities \cite{boffa98,ponson07}. The current
interpretation associates
the value $\zeta\simeq 0.8$ with rupture processes occurring
inside the fracture process zone (FPZ), where elastic interactions
would be screened, and the value $\zeta \simeq 0.45$ with large
scale elastic fracture \cite{ponson06,bonamy06}. 
In two dimensions, the available experimental results, mainly obtained
for paper samples, indicate a roughness exponent in the range $\zeta \simeq 0.6-0.7$ 
\cite{kertesz93,engoy94,salminen03,rosti01}. 

In this work, we investigate the influence of fracture process zone on 
crack roughness in two dimensions through two key variables: material disorder, 
expressed as a distribution in breaking thresholds, and pre-existing notches. 
Material disorder and the size of pre-existing notches play a significant role in 
determining the size of the FPZ ahead of the crack tip. 
When the disorder is weak, the size of the FPZ is small and the material 
fracture response is dictated by the stress concentrations around the notches. On the 
other hand, when the disorder is strong, a relatively large fracture process zone 
is generated ahead of the crack tips. Similarly, the influence of pre-existing notches on FPZ 
in the presence of disorder is non-trivial. This is especially the case when the initial 
notch size is small and disorder is sufficiently strong to allow for significant 
damage accumulation. As the damage starts evolving, 
multiple cracks develop, which in turn influence the stress concentration around the 
initial pre-existing notch.  
Even in the simplest case of non-interacting cracks, the stress fields become 
additive and hence the proportionality with respect to inverse of square root of the initial 
notch size is lost.  The presence of interacting cracks further complicates this scenario 
and the stress concentration around notches depends in a non-trivial fashion on the 
initial notch size. For large notches, the effect of disorder should be weaker since 
the fracture process is dominated by a single crack. 

The question we would like to address is how the roughness of the 
fracture surfaces depends on the material disorder and the relative sizes of 
the pre-existing notches, given their influence on fracture process zone. 
Studies on random fuse model with  uniform and power law disorder have indicated that spatial 
correlations in the damage accumulated prior to the peak load (the maximum load 
before catastrophic failure) are negligible 
and that the damage is accumulated more or less uniformly up to the peak load \cite{jstat1}. 
This suggests that the origin of self-affine roughness in the random fuse model 
should not depend on whether there is strong or weak disorder since the spatial 
correlations are built in the system only at the final stage of macroscopic 
failure. Earlier studies that investigated the effect of disorder on crack roughness 
are controversial: based on two-dimensional disordered beam lattice simulations, 
Ref. \cite{hansenbeam} suggested a universal roughness exponent of $\simeq 0.86$, whereas 
using two-dimensional disordered fuse lattice simulations, 
Ref. \cite{zhang} argued against the universality of roughness exponent. Clearly, 
the situation warrants further investigation especially in light of the role 
played by the FPZ in the current interpretation 
for different values of the roughness exponents. 

\section{Model}
In this paper, we study the effect of disorder and notches on the crack 
roughness, by numerical simulations of the two-dimensional random fuse model (RFM), 
where a lattice of fuses with random threshold are subject to an increasing external voltage
\cite{deArcangelis85,hansen001}. The results show that the roughness exponent 
does not depend on the breaking thresholds disorder strength and on the presence of a notch. 
We consider a triangular lattice of linear size $L$ with a central notch of length
$a_0$. All of the lattice bonds have the same conductance, but the bond breaking
thresholds, $t$, are randomly distributed based on a thresholds
probability distribution, $p(t)$. The burning of a fuse occurs
irreversibly, whenever the electrical current in the fuse exceeds the
breaking threshold current value, $t$, of the fuse. Periodic boundary
conditions are imposed in the horizontal directions ($x$ direction) 
to simulate an infinite system and a constant voltage difference, $V$, is applied
between the top and the bottom of the lattice system bus bars.

A power-law thresholds distribution $p(t)$ is used 
by assigning $t = X^D$, where $X \in [0,1]$ is a uniform 
random variable with density $p_X(X) = 1$ and 
$D$ represents a quantitative measure of disorder. 
The larger $D$ is, the stronger the disorder. This results in 
$t$ values between 0 and 1, with a cumulative
distribution $P(t) = t^{1/D}$. The average breaking 
threshold is $<t> = 1/(D+1)$, and the probability that a fuse 
will have breaking threshold less than the average breaking threshold $<t>$ 
is $P(<t>) = (1/(D+1))^{1/D}$. That is, the larger the $D$ is, the 
smaller the average breaking threshold and the larger the probability 
that a randomly chosen bond will have breaking threshold smaller than the 
average breaking threshold.

Numerically, a unit voltage difference, $V = 1$, is set between the
bus bars (in the $y$ direction) and the Kirchhoff equations are solved to determine the
current flowing in each of the fuses. Subsequently, for each fuse $j$,
the ratio between the current $i_j$ and the breaking threshold $t_j$
is evaluated, and the bond $j_c$ having the largest value,
$\mbox{max}_j \frac{i_j}{t_j}$, is irreversibly removed (burnt).  The
current is redistributed instantaneously after a fuse is burnt
implying that the current relaxation in the lattice system is much
faster than the breaking of a fuse.  Each time a fuse is burnt, it is
necessary to re-calculate the current redistribution in the lattice to
determine the subsequent breaking of a bond.  The process of breaking
of a bond, one at a time, is repeated until the lattice system falls apart. 

Using the algorithm proposed in Ref. \cite{nukalajpamg1}, we have performed
numerical simulation of fracture up to system 
sizes $L = 320$. Our simulations cover an extensive parametric
space of ($L$, $D$ and $a_0$) given by: $L = \{64, 128, 192, 256,
320\}$; $D = \{0.3, 0.4, 0.5, 0.6, 0.75, 1.0\}$; and
$a_0/L = \{1/32, 1/16, 3/32, 1/8, 3/16, 1/4, 5/16, 3/8\}$. A minimum of 200
realizations have been performed for each case, but for many cases
2000 realizations have been used to reduce the statistical error.

\section{Crack Roughness}

\begin{figure*}[hbtp]
\includegraphics[width=16cm]{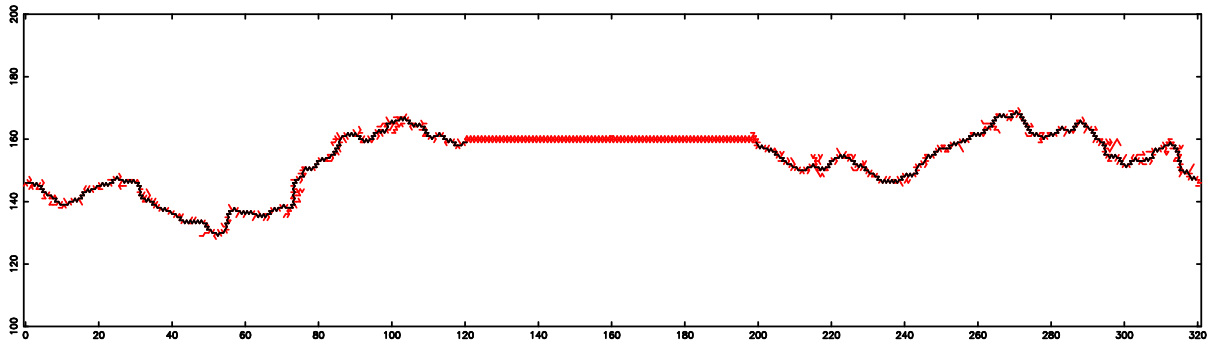}
\caption{(Color online) A typical final crack in a system of size $L = 320$ with initial notch 
size $a_0 = 80$. Note that the crack shows dangling ends and overhangs, which are removed to obtain a 
single valued crack line. The initial central notch is not considered in the roughness calculations.}
\label{fig:crack}
\end{figure*}

Once the sample has failed, we identify the final crack, which typically 
displays dangling ends and overhangs (see Fig. \ref{fig:crack}). We remove them and obtain
a single valued crack line $h_x$, where the values of $x \in [0,L]$. 
For self-affine cracks, the local width, 
$w(l)\equiv \langle \sum_x (h_x- (1/l)\sum_X h_X)^2 \rangle^{1/2}$,
where the sums are restricted to regions of length $l$ and the average
is over different realizations, scales as $w(l) \sim l^\zeta$
for $l \ll L$ and saturates to a value $W=w(L) \sim L^\zeta$ corresponding
to the global width. The power spectrum 
$S(k)\equiv \langle \hat{h}_k \hat{h}_{-k} \rangle/L$, where 
$\hat{h}_k \equiv \sum_x h_x \exp i(2\pi xk/L)$, decays as
$S(k) \sim k^{-(2\zeta+1)}$. When anomalous scaling is present \cite{anomalous,exp-ano,exp-ano2}, 
the exponent describing the system size dependence of the
surface {\it differs} from the local exponent measured for a fixed system
size $L$. In particular, the local width scales as 
$w(\ell) \sim \ell^{\zeta_{loc}}L^{\zeta-\zeta_{loc}}$, so that the global 
roughness $W$ scales as $L^\zeta$ with $\zeta>\zeta_{loc}$. Consequently, the
power spectrum scales as $S(k) \sim k^{-(2\zeta_{loc}+1)}L^{2(\zeta-\zeta_{loc})}$.

\begin{figure}[hbtp]
\includegraphics[width=8cm]{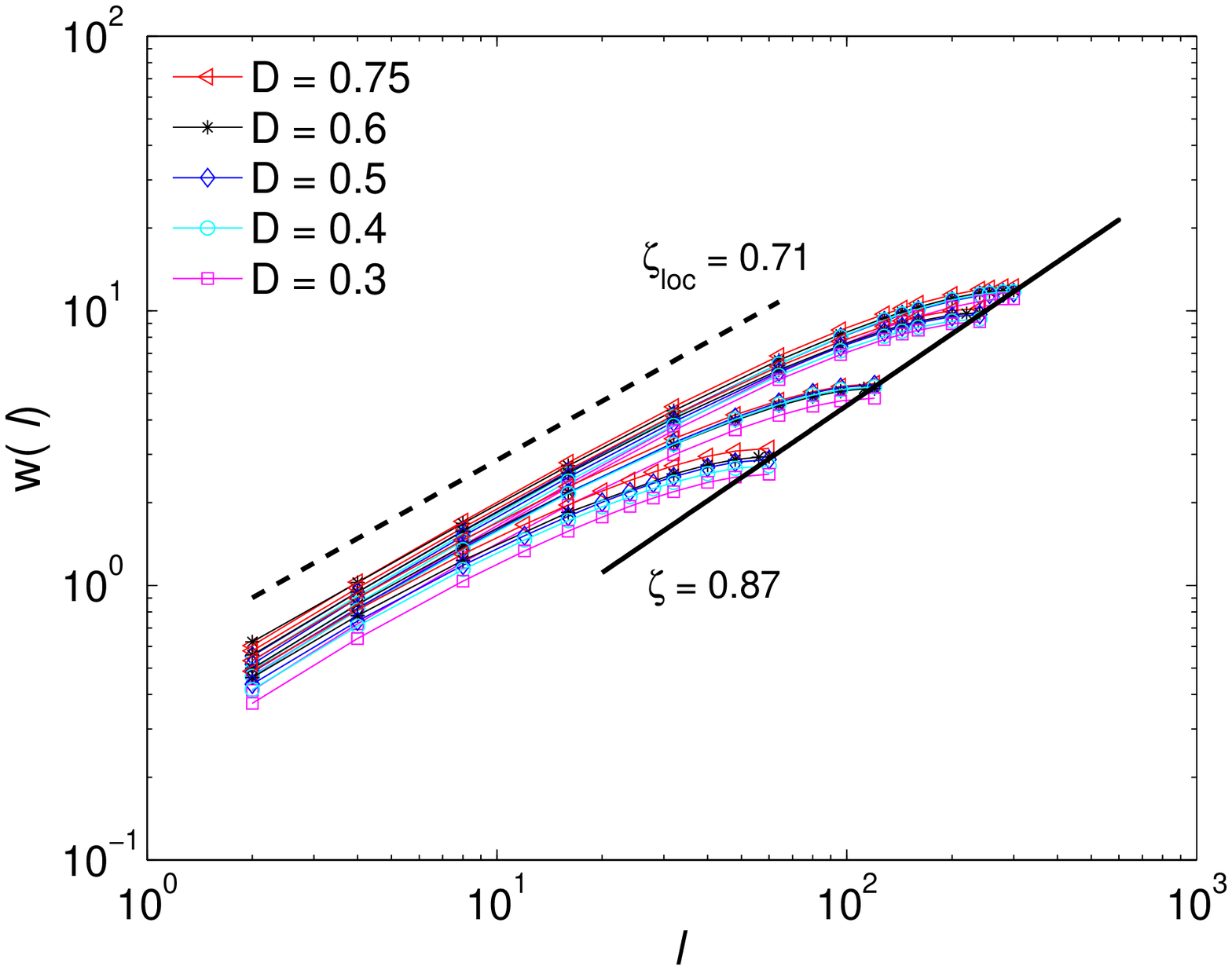}
\includegraphics[width=8cm]{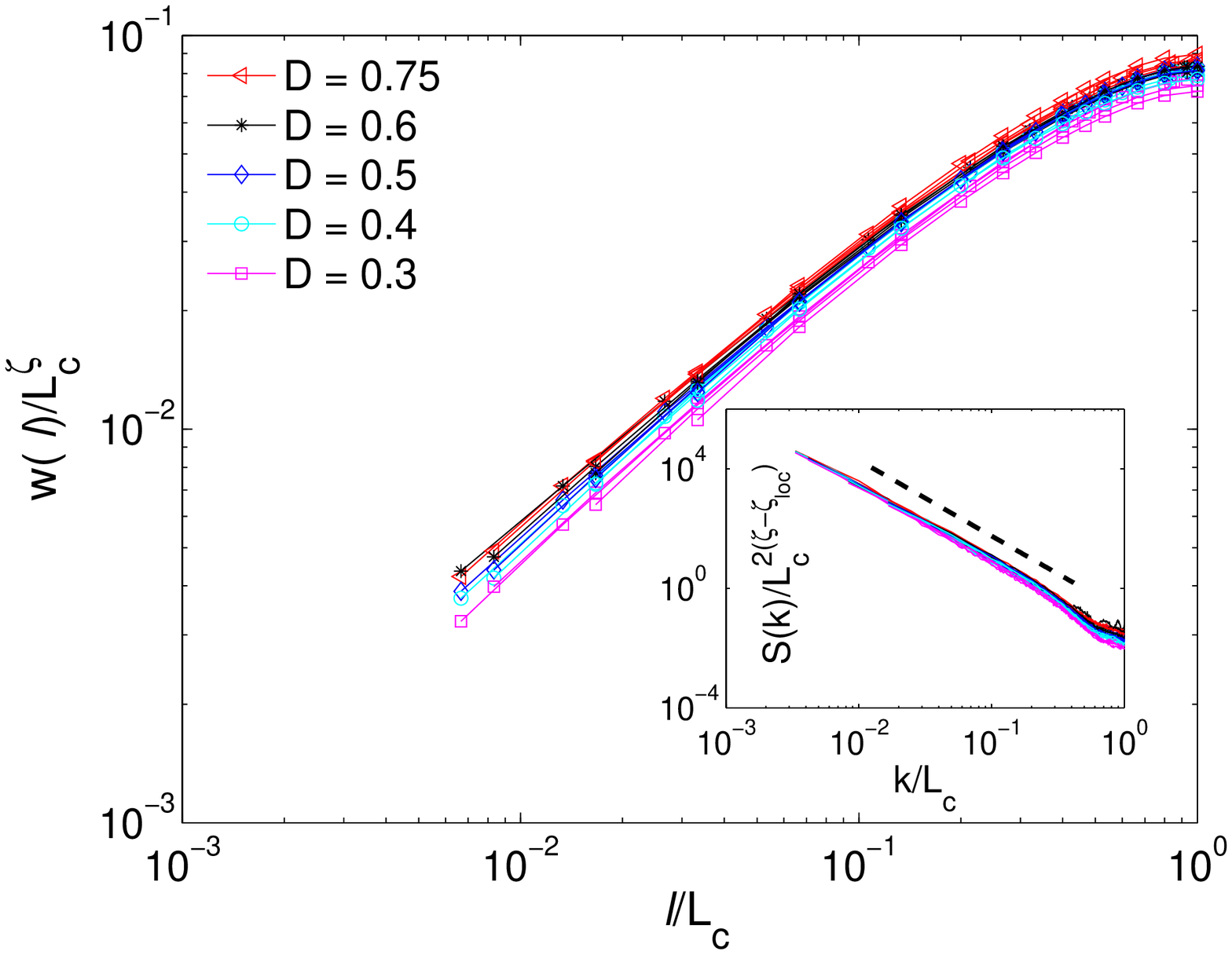}
\caption{(Color online) (a) Scaling of local and global widths $w(l)$ and $W$ of the crack 
for different system sizes $L = \{64,128,256,320\}$, disorder values $D$ and a fixed $a_0/L = 1/16$ value (top). 
The local crack width exponent $\zeta_{loc} = 0.71$ is independent of disorder and differs considerably 
from the global crack width exponent $\zeta = 0.87$. 
(b) Collapse of the crack width data using the anomalous scaling law (bottom). $L_c = (L-a_0)$ is the 
effective length of the crack profile. Collapse of the data 
for a given disorder value implies that local and global roughness exponents are 
independent of disorder. The inset shows collapse of power spectrum $S(k)$ using the anomalous scaling law 
with $\zeta_{loc} = 0.71$ and $\zeta = 0.87$. The slope in the inset defines the local exponent via 
$-(2\zeta_{loc}+1) = -2.48$. (a)-(b) present a total of 20 data sets.}
\label{fig:widthD}
\end{figure}

In the following we investigate the influence of disorder $D$ and 
initial notch size $a_0$ on crack roughness. Figure \ref{fig:widthD}a 
presents the scaling of local and global crack widths in systems with different disorder values 
and an initial relative notch size of $a_0/L = 1/16$. The slopes of the curves presented 
in Fig. \ref{fig:widthD}a suggest that the local roughness exponent 
$\zeta_{loc} = 0.71$ and is independent of the disorder. The global 
roughness exponent is estimated to be $\zeta = 0.87$, and differs 
considerably from the local roughness exponent $\zeta_{loc}$. 
The collapse of the data in Fig. \ref{fig:widthD}b clearly demonstrates that crack widths 
follow anomalous scaling law. The inset in Fig. \ref{fig:widthD}b reports the data collapse of the power spectra 
based on anomalous scaling for different disorder values. 
This collapse of the data once again suggests that local roughness is independent of 
disorder. A fit of the power law decay of the spectrum yields a local 
roughness exponent of $\zeta_{loc}=0.74$. This result is in close agreement 
with the real space estimate and we can attribute the differences to 
the bias associated to the methods employed \cite{sch-95}.

\begin{figure}[hbtp]
\includegraphics[width=8cm]{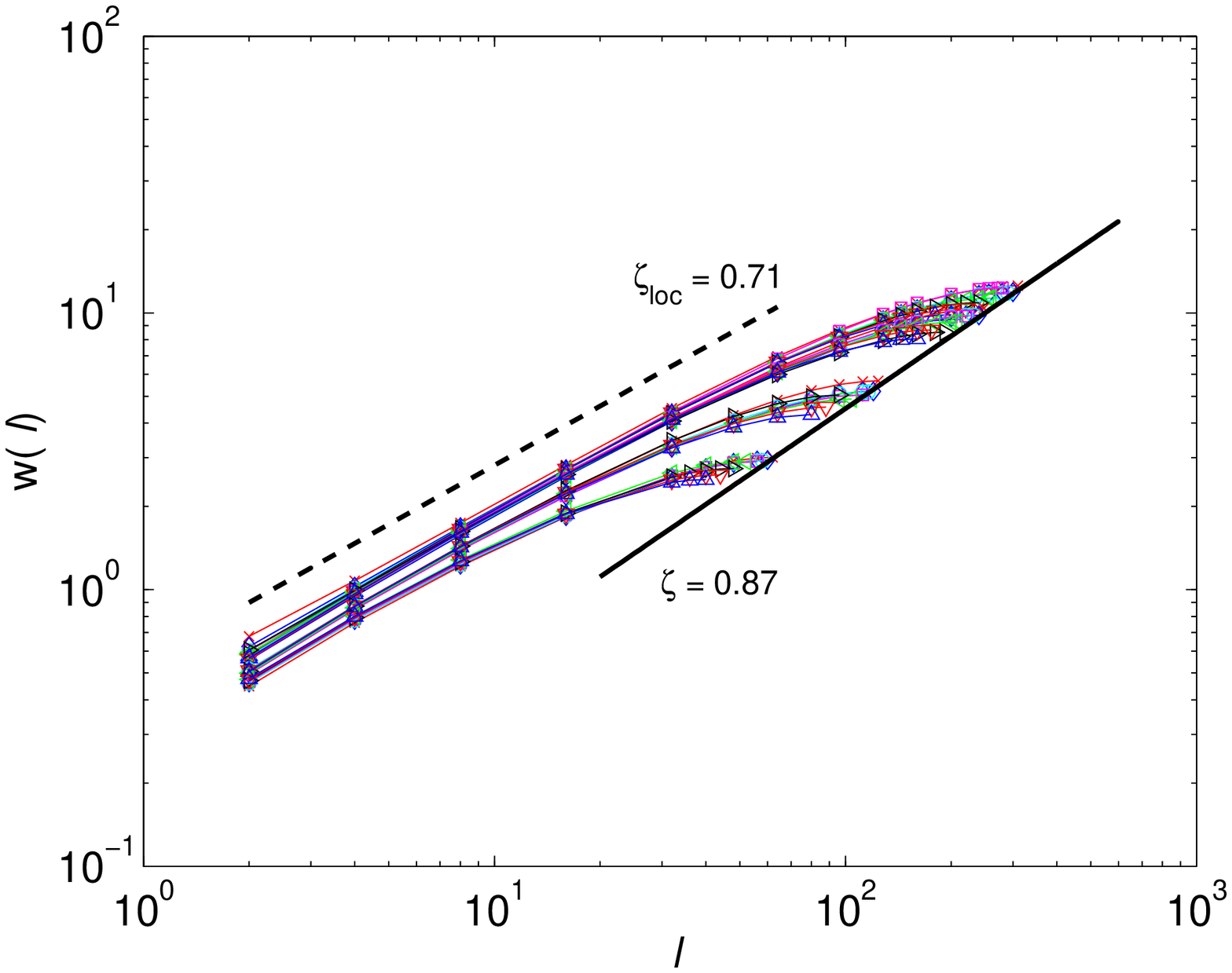}
\includegraphics[width=8cm]{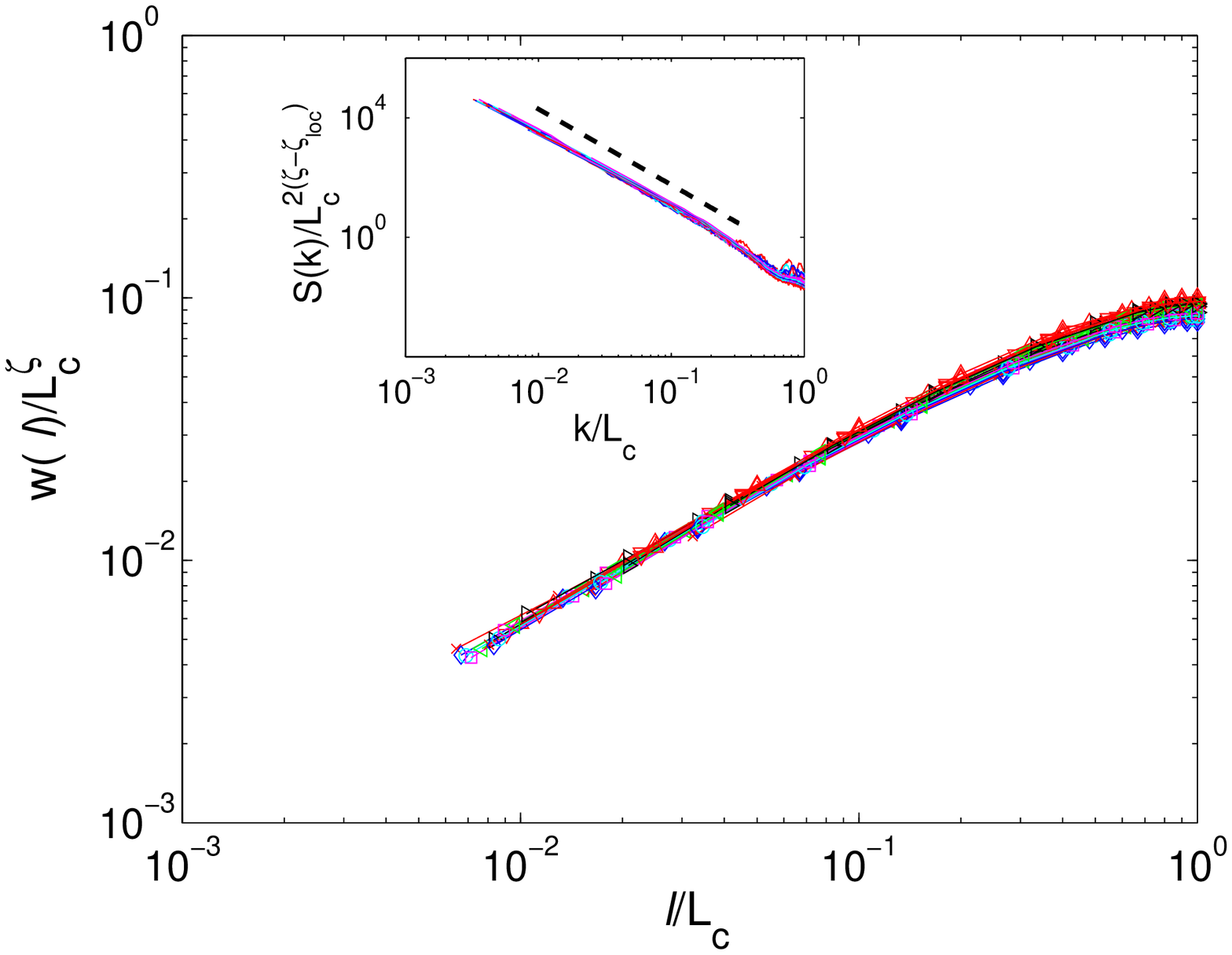}
\caption{(Color online) (a) Scaling of local and global widths $w(l)$ and $W$ of the crack 
for different system sizes $L = \{64,128,256,320\}$, notch sizes $a_0/L = \{1/32, 1/16, 3/32, 1/8, 3/16, 1/4, 5/16, 3/8\}$, 
and a constant disorder of $D = 0.6$ (top). 
Once again, the local crack width exponent $\zeta_{loc} = 0.71$ is independent of notch size $a_0$ 
and differs considerably from the global crack width exponent $\zeta = 0.87$. 
(b) Collapse of the crack width data using the anomalous scaling law (bottom). $L_c = (L-a_0)$ is the 
effective length of the crack profile. The inset shows collapse of power spectrum $S(k)$ using the 
anomalous scaling law with $\zeta_{loc} = 0.71$ and $\zeta = 0.87$. The slope in the inset 
defines the local exponent via $-(2\zeta_{loc}+1) = -2.53$. (a)-(b) 
present a total of 32 data sets.}
\label{fig:widtha0}
\end{figure}

The influence of initial notch size on crack roughness is presented in 
Fig. \ref{fig:widtha0}a. The curves presented in Fig. \ref{fig:widtha0}a represent the 
scaling of local and global widths for various notch sizes $a_0$.  
Once again, the local roughness exponent is estimated to be $\zeta_{loc} = 0.71$ 
and is independent of the initial notch size, whereas the global roughness 
exponent $\zeta = 0.87$. Figure \ref{fig:widtha0}b presents the data collapse of 
crack widths based on anomalous scaling law, which once again confirms that crack 
roughness follows anomalous scaling. 
The collapse of the power spectra in the inset of Fig. \ref{fig:widtha0}b 
for different notch sizes confirms that the 
local roughness is independent of the initial notch size. A fit of the 
power law decay of the spectrum yields a local roughness exponent value 
of $\zeta_{loc}=0.77$. The close agreement of these results with the 
$\zeta_{loc} = 0.72$ obtained for the unnotched, strong disorder case \cite{zapperi05}
indicates that the crack roughness is universal and is independent of 
disorder and initial notch size. The global crack width $W$ however scales as 
$W \equiv (\langle (h_b-\bar{h_b})^2\rangle)^{1/2} \sim L^\zeta$ with $\zeta = 0.87\pm0.03$, 
and is also independent of disorder and crack size.

%\begin{figure}[hbtp]
%\includegraphics[width=8cm]{halfseg_notch_L320_D0_60_vara0_inset.eps}
%\caption{(Color online) (The local width $w(l)$ of the crack 
%for different notch sizes $a_0$ and a constant disorder of $D = 0.6$.
%The crack width scaling is independent of notch size, and exhibits a local roughness 
%exponent of $\zeta_{loc} = 0.71$. Power spectrum of the crack $S(k)$ is 
%shown in the inset. The slope defines the local exponent via 
%$-(2\zeta_{loc}+1) = -2.53$.}
%\label{fig:widthD}
%\end{figure}

The scaling properties of the crack profiles $h(x)$ can also be 
studied using the probability density distribution $p(\Delta h(\ell))$ of the height 
differences $\Delta h(\ell) = [h(x+\ell) - h(x)]$ of the crack profile between any 
two points on the reference line ($x$-axis) separated by a distance $\ell$. Recently, there 
has been a debate over the scaling of this $p(\Delta h(\ell))$ distribution \cite{procaccia,jstat2,santucci}, i.e., 
whether the scaling properties of $p(\Delta h(\ell))$ can be described by a single scaling exponent $\zeta$ 
or multiple scaling exponents are required to describe the scaling of $p(\Delta h(\ell))$. The self-affine 
property of the crack profiles implies that the probability density distribution $p(\Delta h(\ell))$ 
follows the relation 
\begin{eqnarray}
p(\Delta h(\ell)) & \sim & \ell^{-\zeta_{loc}} f\left(\frac{\Delta h(\ell)}{\ell^{\zeta_{loc}}}\right) \label{pdelt}
\end{eqnarray}
whereas multiscaling of fracture surface implies that Eq. \ref{pdelt} is not valid. 
Although multiscaling of $p(\Delta h(\ell))$ was argued in 
Ref. \cite{procaccia}, it has been shown in references \cite{jstat2,santucci} that $p(\Delta h(\ell))$ 
follows a self-affine monoscaling relation given by Eq. \ref{pdelt} and that multiscaling is an artefact 
that results at small scales due to the removal of crack profile overhangs. In the following, 
we investigate whether disorder has any influence on the scaling of $p(\Delta h(\ell))$. 

\begin{figure}[hbtp]
\includegraphics[width=8cm]{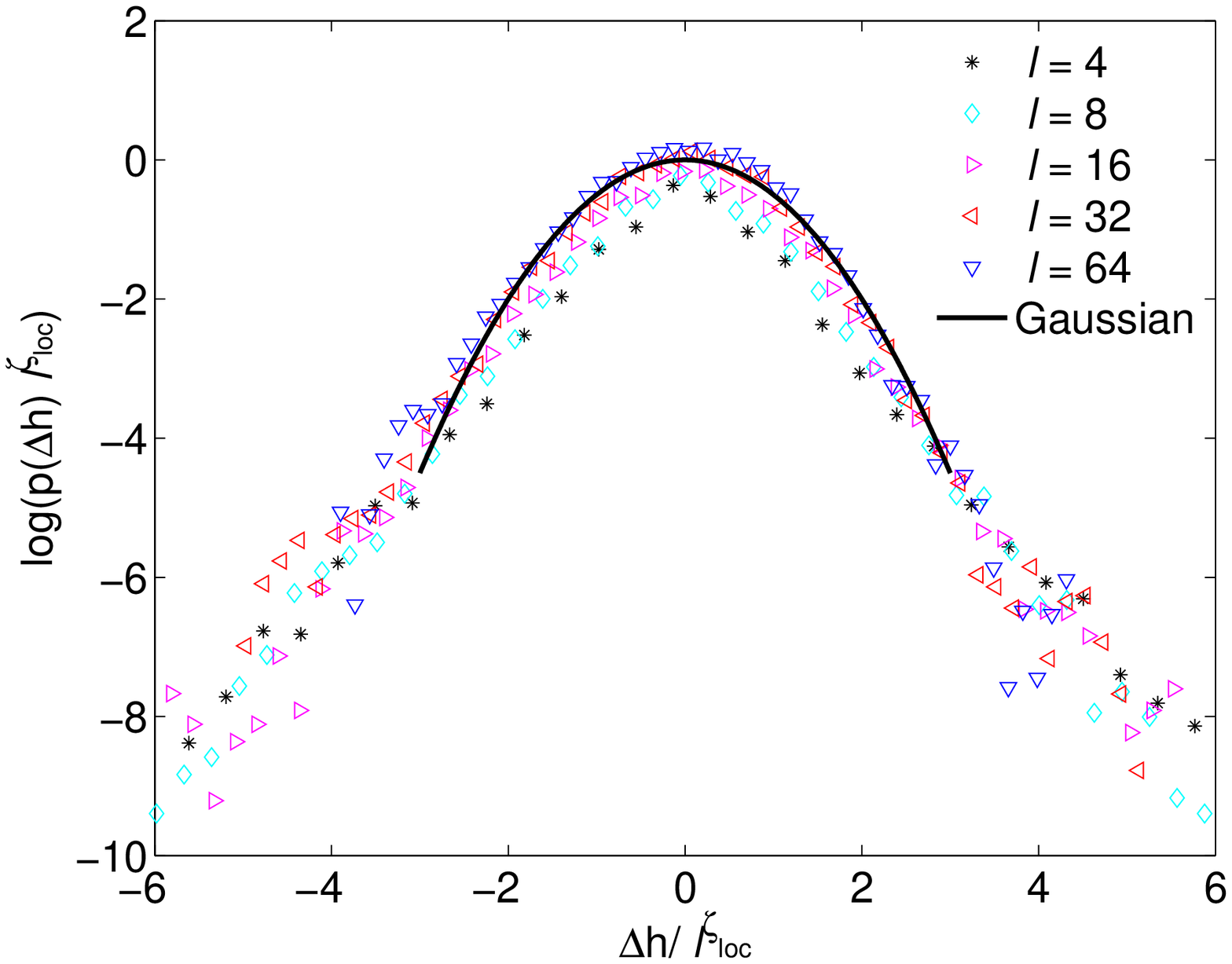}
\caption{(Color online) The logarithm of probability distributions $p(\Delta h(\ell))$ 
of the height differences $\Delta h(\ell) = [h(x+\ell) - h(x)]$ of the crack profile 
$h(x)$ for various bin sizes $\ell = 4, 8, 16, 32, 64$. The results are for crack profiles 
obtained using a system size of $L = 320$, a disorder of $D = 0.75$ and a relative 
crack size of $a_0/L = 1/16$. As a guide for the eye, we present a Gaussian fit for 
$\ell = 32$. It can be seen that the central parts of the distributions are Gaussian, 
but the tails do not follow a Gaussian.}
\label{fig:pdelth}
\end{figure}

First, we present the probability distributions $p(\Delta h(\ell))$ 
of the height differences $\Delta h(\ell) = [h(x+\ell) - h(x)]$ for various bin sizes 
$\ell = 4, 8, 16, 32, 64$ for a disorder of $D = 0.75$, a system size of $L = 320$, and a 
relative crack size of $a_0/L = 1/16$. Figure \ref{fig:pdelth} shows the 
collapse of the central parts of the probability distributions $p(\Delta h(\ell))$ for larger 
$\ell$ values, but still within the local width scaling regime. The deviation 
for smaller $\ell$ values may be attributed to steps in the single-valued crack height profiles, 
which inevitably arise due to the removal of overhangs on the crack surface. For larger $\ell$ values, 
the central parts of these distributions approach Gaussian, 
but clear deviations can be observed in the tails of the distribution from a Gaussian distribution.   
Similar deviations in the tails were observed for the 
uniform disorder case as well \cite{jstat2,santucci}. 

\begin{figure}[hbtp]
\includegraphics[width=8cm]{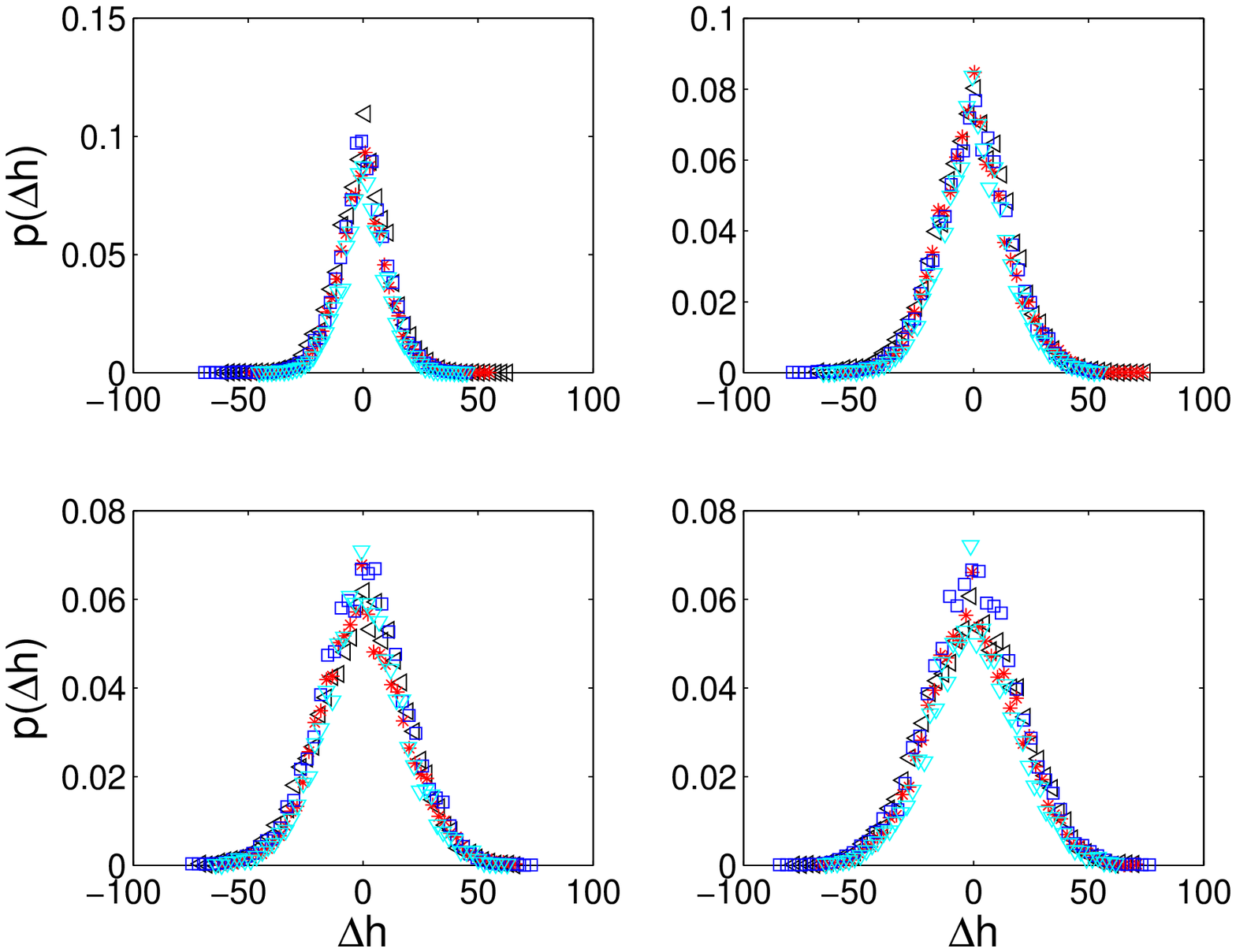}
\caption{(Color online) Collapse of the probability density distributions $p(\Delta h(\ell))$ 
of the crack profiles for various disorder values $D = 0.3, 0.4, 0.5, 0.75$. (a) $\ell = 32$ (top left), 
(b) $\ell = 64$ (top right), (c) $\ell = 96$ (bottom left), and (d) $\ell = 128$ (bottom right). 
The results are obtained for a system size of $L = 320$.}
\label{fig:Dpdelt}
\end{figure}

Second, we present the collapse of $p(\Delta h(\ell))$ distributions in Figs. \ref{fig:Dpdelt}(a)-(d) 
for various bin sizes $\ell$ and disorders. The collapse of these $p(\Delta h(\ell))$ distributions 
for various disorders $D$ at each bin size $\ell$ indicates 
that $p(\Delta h(\ell))$ distributions and the roughness exponent $\zeta_{loc}$ are 
unaffected by the material disorder. 

\section{Discussion}

\begin{figure}[hbtp]
\begin{tabular}{ccccc}
\includegraphics[width=4cm]{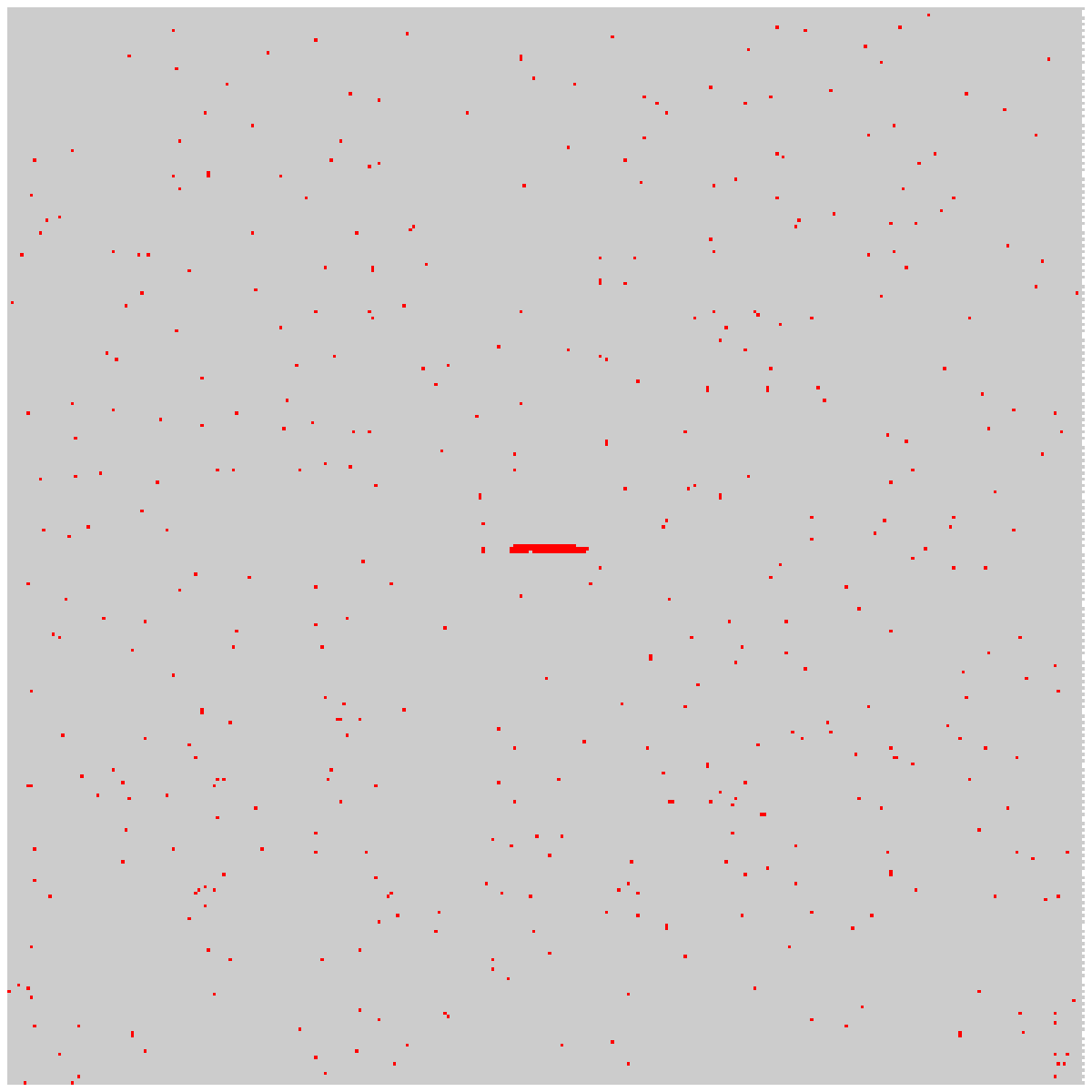} & 
\includegraphics[width=4cm]{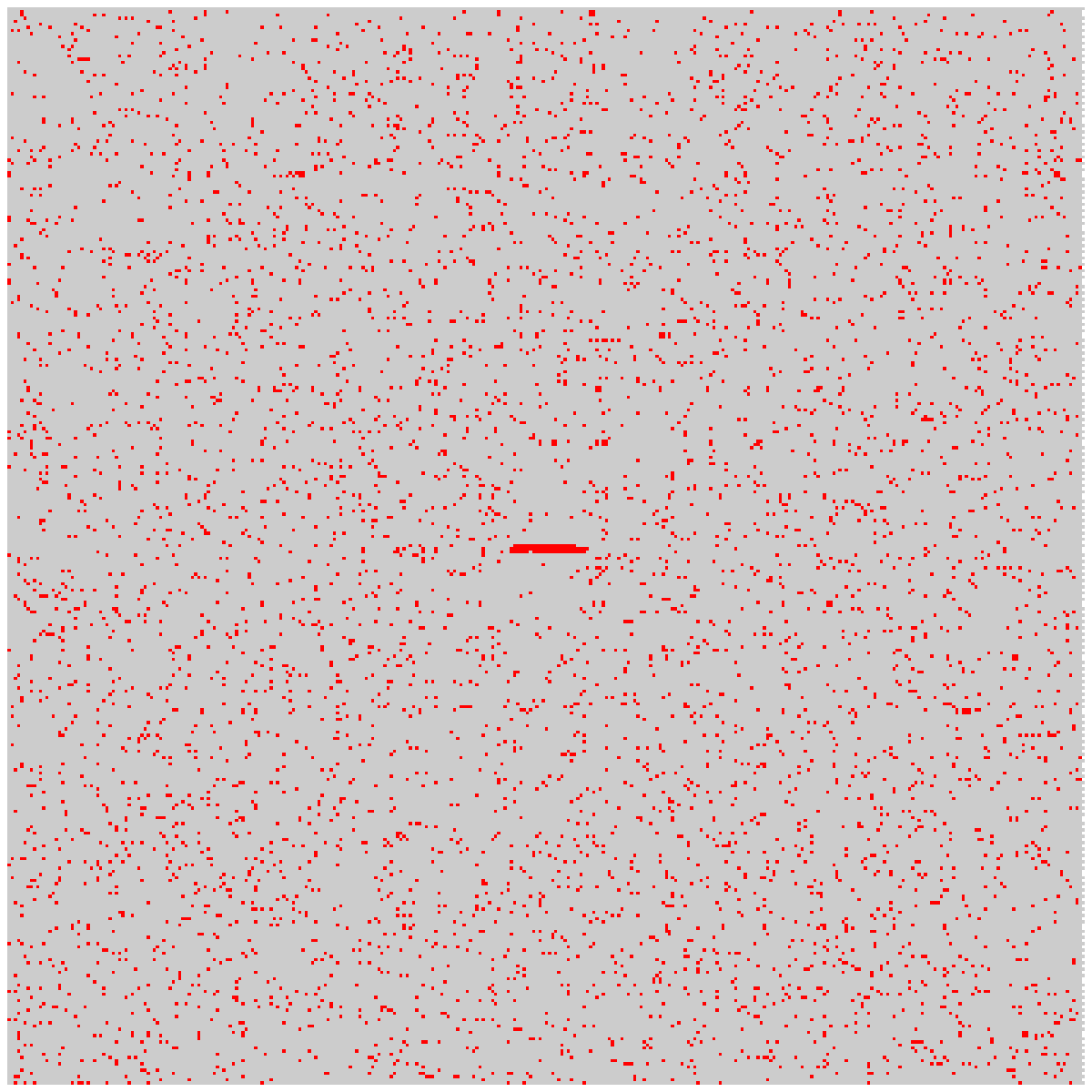} \\
\includegraphics[width=4cm]{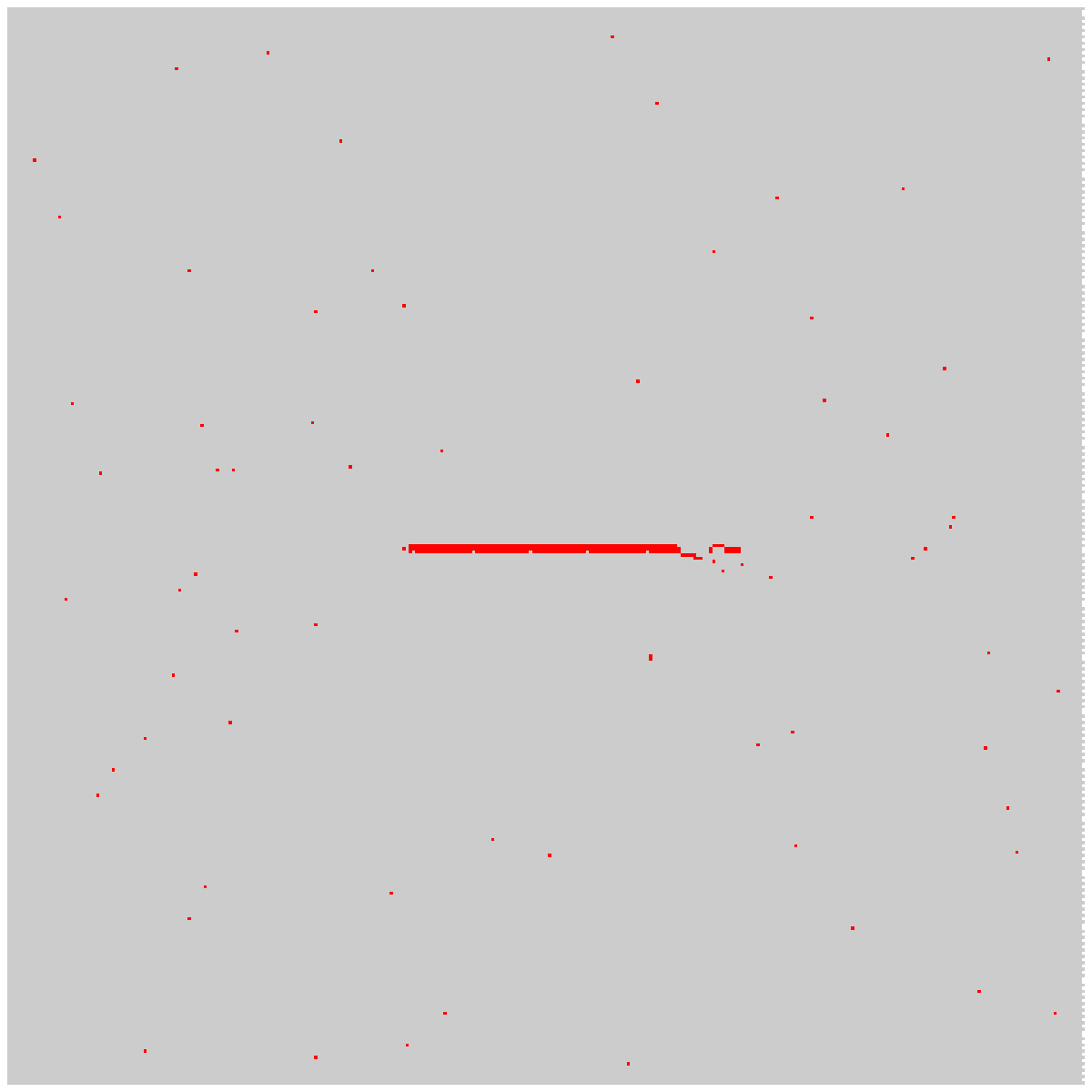} &   
\includegraphics[width=4cm]{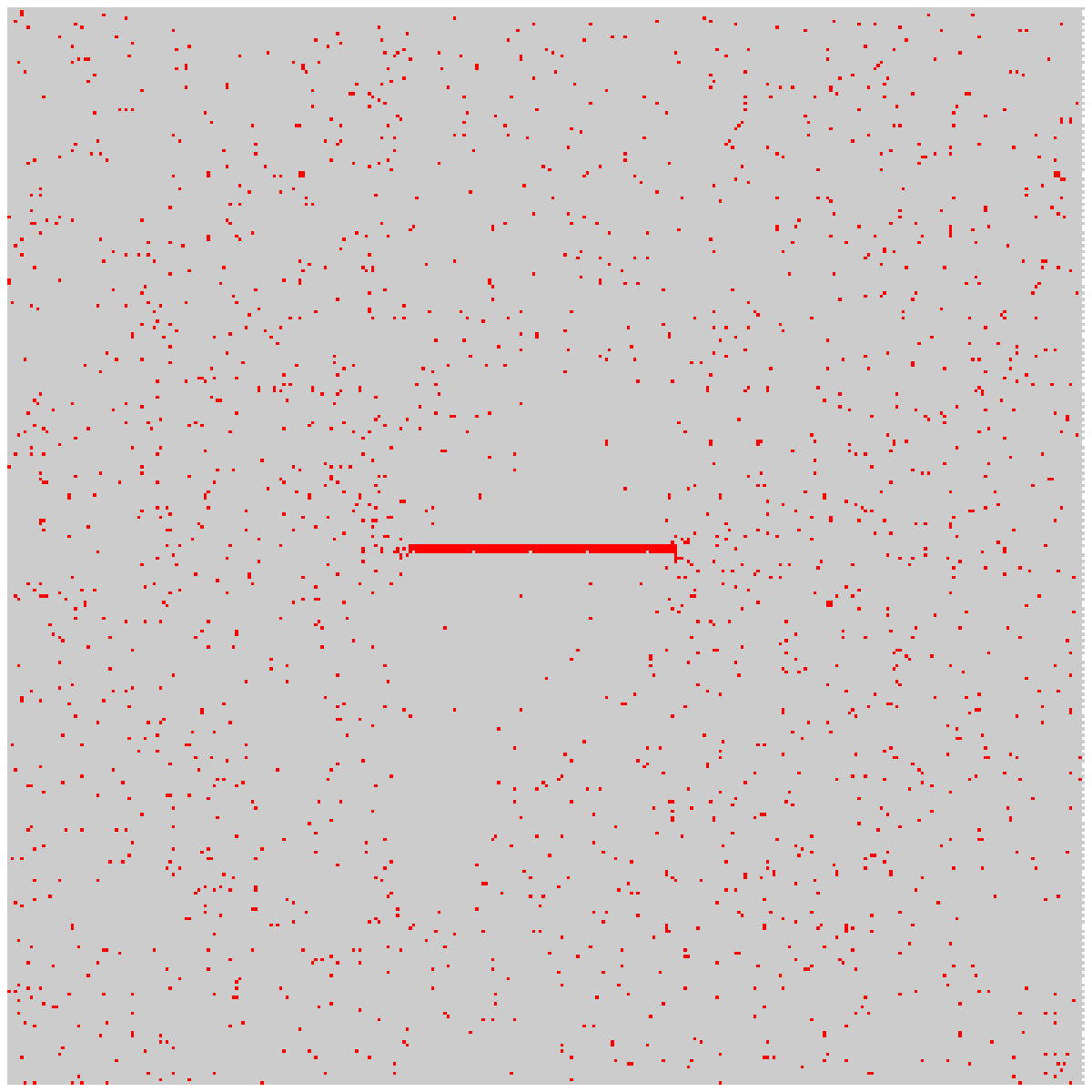} \\  
\end{tabular}
\caption{(Color online) Snapshots of damage and crack evolution in disordered notched 
specimens of size $L = 320$. (a) $D = 0.3$ and $a_0 = 20$ (top left); (b) $D = 0.6$ and $a_0 = 20$ (top right);  
(c) $D = 0.3$ and $a_0 = 80$ (bottom left); (d) $D = 0.6$ and $a_0 = 80$ (bottom right).}
\label{fig:complex}
\end{figure}

In summary, the evidence presented in this paper indicates that the crack 
surface roughness is unaffected by the material disorder and the presence of 
pre-existing notches. This can be inferred from the fact that 
material disorder, whether strong or weak, has a significant influence on the 
amount of damage accumulated prior to the peak load; however, the spatial  
correlations in the damage accumulated prior to the peak load are negligible \cite{jstat1}. 
Indeed, Figs. \ref{fig:complex}(a)-(d) show the snapshots of damage and crack 
propagation at peak load in typical fracture simulations of system size $L = 320$ 
having weak to strong disorder and small to large pre-existing notches. 
By following the damage growth process, one can easily see from these figures that 
there is very little crack extension at peak load whether the 
material is strongly disordered or weakly disordered. 
In addition, Fig. \ref{fig:complex} shows that the FPZ can not be defined from a 
single damage snapshot, but it is necessary to  average the damage over many realizations of the disorder. 
When this is done, we find that $\xi_{FPZ}$ depends strongly on disorder $D$ (ranging
from one to $12$ lattice units for the values considered) 
but only weakly on $a_0$ and $L$ \cite{alavanew}. The independence of the roughness
exponent on $D$ suggests that self-affinity in the random fuse model 
is  not related to the FPZ. 

%\begin{figure*}[hbtp]
%\end{figure*}

This is further corroborated by our numerical simulations on a simplified random fuse model 
in which failure events form a connected crack thereby excluding
damage nucleation in the bulk \cite{jstat2}. In this model, after breaking the weakest fuse,
successive failure events are only allowed on fuses that are connected
to the crack. Otherwise, the rules of this simplified model strictly follow
those of the usual RFM. Consequently, this model tracks only the connected crack 
along with its dangling ends in a disordered medium, and hence forms the most simplified 
model to study the effect of disorder on crack roughness. As was shown in Ref. \cite{jstat2}, this simplified model exhibits the same roughness and height-height correlation characteristics as that 
of conventional RFM. Even for this simplified model, the collapse of the power spectrum and 
local width of crack profiles for different disorders ($D = \{0.25,0.4,0.5,0.6,0.75,1.0\}$) results in a roughness 
exponent value of $\zeta_{loc} = 0.7$, which clearly 
demonstrates once again that disorder is irrelevant for identifying crack roughness exponent.

%\begin{figure}[hbtp]
%\includegraphics[width=8cm]{single_notch_L320_inset.eps}
%\caption{(Color online) Collapse of the power spectrum $S(k)$ of the connected crack 
%in the simplified random fuse model for different disorders. 
%The slope defines the local exponent via $-(2\zeta_{loc}+1) = -2.47$. Inset shows the 
%the collapse of local width $w(l)$ of the crack 
%for different disorders. Once again, a roughness exponent of 0.70 is obtained.}
%\label{fig:sing}
%\end{figure}

As also shown in Ref. \cite{jstat2}, when the branching of the cracks or damage within 
the fracture process zone is not allowed thereby limiting the crack extension to only the 
crack tips, a local roughness exponent of $0.5$ is obtained. But, as soon as branching was present, the
value of the roughness exponent was increased from $\zeta_{loc} = 0.5$ to $\zeta_{loc} \simeq 0.7$. 
This may also explain why there is no transition in the value of the roughness exponent from strong to weak
disorder in the presented simulations since damage is always present even for the lowest values of 
disorders considered. A similar conclusion was reached by Bouchbiner et al. studying a model
for crack growth with damage nucleation \cite{procaccia2}. According to this work, as soon as a FPZ
was introduced in the model the roughness exponent increases from $\zeta_{loc} = 0.5$ to a higher
value. Thus it appears that the roughness exponent in two dimensions does not depend on the size
of the FPZ, but only on the fact that a FPZ is present or not. In three dimensions the situation
should be different since experiments suggest that the roughness exponent displays a crossover precisely 
at the FPZ size. It would be interesting to investigate this issue using three dimensional simulations.

\par
\vskip 1.00em%
\noindent
{\bf Acknowledgment} \\ This research is sponsored by the
Mathematical, Information and Computational Sciences Division, Office
of Advanced Scientific Computing Research, U.S. Department of Energy
under contract number DE-AC05-00OR22725 with UT-Battelle, LLC. MJA and SZ gratefully
thank the financial support of the European Commissions
NEST Pathfinder programme TRIGS under contract NEST-2005-PATH-COM-043386. 
MJA also acknowledges the financial support from 
The Center of Excellence program of the Academy of Finland.


\begin{thebibliography}{99}

\bibitem{breakdown} H. J. Herrmann and S. Roux (eds.), {\em
        Statistical Models for the Fracture of Disordered Media},
        (North-Holland, Amsterdam, 1990). 

\bibitem{alava06}
	M. J. Alava, P. K. V. V. Nukala, and S. Zapperi, 
	Advances in Physics {\bf 55}, 349 (2006).

\bibitem{man}
        B. B. Mandelbrot, D. E. Passoja, and A. J. Paullay,
        Nature (London) {\bf 308}, 721 (1984).

\bibitem{bouch}
	For a review see E. Bouchaud, J Phys. Condens. Matter {\bf 9}, 4319 (1997).
	E. Bouchaud, Surf. Rev. Lett. {\bf 10}, 73 (2003).
\bibitem{metals}
K.J. Maloy, A. Hansen, E.L. Hinrichsen, and S. Roux, Phys. Rev. Lett. 68, 213 (1992);
E. Bouchaud, G. Lapasset, J. Plan{\'e}s, and S. Nav{\'e}os, Phys. Rev. B 48, 2917 (1993).
\bibitem{glass}
  P. Daguier, B. Nghiem, E. Bouchaud, and F. Creuzet, Phys. Rev. Lett. 78, 1062 (1997).
\bibitem{rocks}
J. Schmittbuhl, S. Roux, and Y. Berthaud, Europhys. Lett. 28, 585 (1994).
J. Schmittbuhl, F. Schmitt, and C. Scholz, J. Geophys. Res. 100, 5953 (1995).
\bibitem{cera}
J.J. Mecholsky, D.E. Passoja, and K.S. Feinberg-Ringel,
 J. Am. Ceram. Soc. 72, 60 (1989).

\bibitem{ponson06}
	L. Ponson, D. Bonamy, and E. Bouchaud, 
	Phys. Rev. Lett. {\bf 96}, 035506 (2006).

\bibitem{boffa98}
	J. M. Boffa, C. Allain, and J. Hulin,
	Eur. Phys. J. A {\bf 2}, 281 (1998).

\bibitem{ponson07}
L. Ponson, H. Auradou, M. Pessel, V. Lazarus, and J.-P. Hulin,
preprint arXiv:0704.2925

\bibitem{bonamy06}
	D. Bonamy, L. Ponson, S. Prades, E. Bouchaud, and C. Guillot, 
	Phys. Rev. Lett. {\bf 97}, 135504 (2006).

\bibitem{kertesz93}
Kert\'esz J, Horvath V~K and Weber F 1993 {\em Fractals\/} {\bf 1} 67

\bibitem{engoy94}
Engoy T, Maloy K~J, Hansen A and Roux S 1994 {\em Physical Review Letters\/}
  {\bf 73} 834

\bibitem{salminen03}
Salminen L~I, Alava M~J and Niskanen K~J 2003 {\em Eur. Phys. J. B\/} {\bf 32}
  369

\bibitem{rosti01}
Rosti J, Salminen L~I, Sepp\"al\"a E~T, Alava M~J and Niskanen K~J 2001 {\em
  Eur. Phys. J. B\/} {\bf 19} 259

\bibitem{jstat1}
 P. K. V. V. Nukala, S. Simunovic, and S. Zapperi,
 J. Stat. Mech.: Theor. Exp.  P08001 (2004).


\bibitem{hansenbeam}
	B. Skjetne, T. Helle, and A. Hansen, 
	Phys. Rev. Lett. {\bf 87}, 125503 (2001).

\bibitem{zhang}
	X. Zhang, M. A. Knackstedt, D. Y. C. Chan, and L. Paterson, 
	Europhys. Lett. {\bf 34}, 121-126 (1996)

\bibitem{deArcangelis85}
	L. de Arcangelis, S. Redner, and H. J. Herrmann,
	J. Phys. (Paris) Lett. {\bf 46} 585 (1985).
	
\bibitem{hansen001}
	A. Hansen and S. Roux, 
	{\em Statistical toolbox for damage and fracture}, 17-101, in book 
	{\em Damage and Fracture of Disordered Materials}, eds.  
	D. Krajcinovic and van Mier, Springer Verlag, New York, 2000. 

\bibitem{nukalajpamg1}
	P. K. V. V. Nukala, and S. Simunovic,
	J. Phys. A: Math. Gen. {\bf 36}, 11403 (2003).


\bibitem{anomalous}
J. M. L{\'o}pez, M. A. Rodr{\'i}guez, and R. Cuerno, Phys. Rev. E 56, 3993 (1997). 
\bibitem{exp-ano}
J. M. L{\'o}pez and J. Schmittbuhl, Phys. Rev. E 57, 6405 (1998).
\bibitem{exp-ano2}
S. Morel, J. Schmittbuhl, J. M. L{\'o}pez, and G. Valentin, Phys. Rev. E 58, 6999 (1998).

\bibitem{sch-95}
	J. Schmittbuhl, J. P. Vilotte, and S. Roux, 
	Phys. Rev. E {\bf 51}, 131-147 (1995). 

\bibitem{zapperi05}
	S. Zapperi, P. K. V. V. Nukala, and S. Simunovic,
	Phys. Rev. E {\bf 71}, 026106 (2005). 

\bibitem{procaccia}
	E. Bouchbinder, I. Procaccia, and S. Sela,
	J. Stat. Phys. {\bf 125}, 1029 (2006). 

\bibitem{jstat2}
	M. J. Alava, P. K. V. V. Nukala, and S. Zapperi,
	J. Stat. Mech.: Theor. Exp.  L10002 (2006).

\bibitem{santucci}
	S. Santucci, K. J. Maloy, A. Delaplace, 
	J. Mathiesen, A. Hansen, J. O. H. Bakke, 
	J. Schmittbuhl, L. Vanel, and P. Ray, 
	Phys. Rev. E {\bf 75}, 016104 (2007).

\bibitem{alavanew}
M. J. Alava, P. K. V. V. Nukala, and S. Zapperi,
preprint (2007).


\bibitem{procaccia2}
E. Bouchbinder, J. Mathiesen,  and I. Procaccia,
Phys. Rev. Lett. 92, 245505 (2004) 


\end{thebibliography}
\end{document}